\documentstyle[aps,twocolumn,epsf]{revtex}

\begin{document}
\title{Motion of grains, droplets, and bubbles in fluid-filled nano-pores}
\author{Nazar Sushko and Marek Cieplak}
\address{Institute of Physics, Polish Academy of Sciences, Al. Lotnikow 32/46,
02-668 Warsaw, Poland}
\address{
\centering{
\medskip\em
{}~\\
\begin{minipage}{14cm}
Molecular dynamics studies of nono-sized rigid grains, droplets and bubbles
in nano-sized pores indicate that the drag force may have a hydrodynamic
form if the moving object is dense and small compared to the pore diameter.
Otherwise, the behavior is non-hydrodynamic. 
The terminal speed is insensitive to whether the falling droplet
is made of liquid or a solid.
The velocity profiles within droplets and bubbles
that move in the pore are usually non-parabolic and distinct
from those corresponding to individual fluids.
The density profiles indicate motional shape distortion of 
the moving objects.
{}~\\
{}~\\
\end{minipage}
}}
\maketitle

\section{INTRODUCTION.}

Classical problems of motion of objects through
fluids continue to be of interest because
of their multiple applications in natural sciences and because of
their frequent relation to fundamental issues.
There is a variety of possible physical situations.
The moving objects may may be rigid or fluid. They may
also span many length scales:
from macroscopic, such as snowflakes \cite{Kajikawa}
or sediments \cite{Allen}, to micron-sized, such as water
droplets and aerosols in clouds \cite{Pruppacher,Seinfeld}.
A new frontier arises in the context of micro- and nano-scale 
machinery \cite{Kim,Ho} which sometimes involves flows through
narrow pores. Two-phase flows in micro- and nano-pores
would require understanding of flows and
interactions between nano-sized droplets. One recent 
example of this situation is provided by the surface force
apparatus studies of capillary condensation in a nanoscale
pore \cite{Kohonen}. Droplets,
bubbles, and grains 
of various sizes
may move at various characteristic speeds.
At large Reynolds numbers, smooth trajectories may become
replaced by various classes of complex trajectories, such as
observed in motion of falling disks \cite{Nori} or 
bubbles \cite{Kelley,Krishna}. Another complexity may appear 
due to the droplet/bubble shape deformation, 
possibly combined with splitting and stacking
\cite{Maxworthy,Moore} or due to a bubble collapse. 

In this paper, we focus on issues arising at the
molecular level aspects of such problems by considering nano-sized
objects moving in larger but still nano-sized pores at low Reynolds numbers.
Our interest is in probing physics of nano-scale drag phenomena
in simple molecular models and in assessing the validity of continuum
physics concepts at this length scale.
Our main purpose is to compare motion of droplets and bubbles
to that of the solid spheres.
The spheres are built of fluid-like atoms that
are tethered to amorphously located centers.
Our studies are based on the molecular dynamics (MD) 
approach \cite{Haile,Tildesley,cicc,Koplik,Computers} which allows us to
monitor motion of individual atoms but its use is restricted to nano-second
time scales. Such atomic models should apply literarily only to the
microscopic-scale systems but they may also be considered
as toy models of larger droplets. Note also, that rupture or
coalescence of droplets \cite{Koplik,Koplik1,Koplik2,Cieplak},
no matter how big, necessarily comes to a stage in which the
interface morphology involves subcontinuum physics of nano-scale tendrils. 

\begin{figure}
\epsfxsize=3.7 in
\centerline{\epsffile{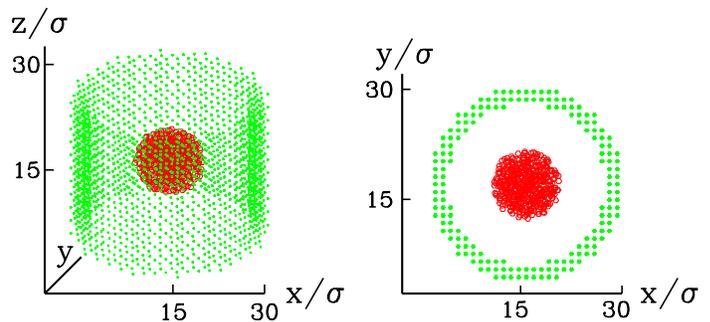}}
\vspace*{-2cm}
\caption{A snapshot of solid sphere moving through a fluid filled
cylindrical pore. Only the solid and wall atoms 
are shown for clarity. The solid consists of 454 tethered atoms 
that are tethered to amorphously placed centers.
There are 2212 pore wall atoms which are fixed rigidly. 
The space between the solid and the pore is filled with 2281 fluid atoms
(the case of fluid B).
The left-hand figure
represents a stereographic projection whereas the right-hand figure
represents a projection from the top.} 
\label{wall}
\end{figure}

A natural way to study motion
of an object in a fluid is to apply a driving force
and determine a drag force, $F_d$,
that brings the motion to a stationary state.
The simulations on
hard sphere systems \cite{Alder,Alder1,Vergeles} have 
indicated validity of
Stokes's law \cite{Stokes},
\begin{equation}
F_d=6\pi R\eta v,\label{'stokes'} \;\;,
\end{equation}
down to atom sized objects, when average quantities are monitored. 
In the Stokes law above, 
$R, \eta$, and $v$ denote, respectively, the radius of the moving sphere, 
the viscosity of the surrounding fluid, and
the velocity of the body relative to an undisturbed fluid.
However, the continuum mechanics predictions \cite{Brenner}
break down and lead to divergent forces
when one considers an approach  of a sphere
to a fixed wall at a constant velocity. 
The MD results \cite{Vergeles}, on the other hand, 
give evidence for a scenario in which the fluid
atoms escape the region that is squeezed between the sphere and the wall,
preventing any buildup in the force.

This paper provides a pilot MD study of the drag force phenomena
that relate to the motion of droplets and bubbles. This study is
aimed at making a comparison to a similar MD analysis of solid
nanoscale objects performed by Vergeles et al. \cite{Vergeles}.
We demonstrate that the drag force on droplets is similar to that
on solid objects but the droplets undergo motional distortion.
The MD results for the motion of droplets essentially agree with
hydrodynamic predictions but this is not so in the case of
the bubbles. The bubbles that we study have densities which are
so low that they correspond to the crossover region between
the sub-continuum and continuum physics \cite{Knudsen} and a 
non-hydrodynamic behavior, at such small length scales, is
not surprising. Another interesting finding is that the 
surface tension effects at nano-scale confinement generally
yield velocity profiles such that the velocity in the
single fluid regions is quite distinct from that corresponding
to one fluid flows.

We start our analysis in
Section 2 where construction of the nanoscale pore is described
and properties of the fluid that fills it are established.
The overall geometry is illustrated in figure \ref{wall} and the
$z$-direction corresponds to the pore axis.
The primary role of the pore is to pin the fluid to its walls and thus
to provide a stationary environment in which motion of falling
objects can be studied. 
Our objective in this paper is focused on comparing various drag force
phenomena at nanoscale length scale, as discussed within a 
simple molecular model of a generic nature,
and not the boundary phenomena arising near the confining wall.
We take then the wall to be built of Lennard-Jones atoms fixed rigidly
on an fcc lattice and adjust parameters that minimimize layering
in the density profile near the wall. The reason is that it seems
proper to consider first the simplest kind of the drag phenomena
in which special wall-related effects are missing.
Studies of near-the wall physics and of excitations
in the solid due to a flow would require considering a more realistic
model, such as used in studies of fluid flows in carbon
nanotubes by Tuzun et al. \cite{Tuzun}.

We consider two kinds of fluids: one at a dense liquid
density (A) and another at a dense gas density (B). Both are described 
by the Lennard-Jones potential
\begin{equation}
u_{LJ}(r)=4\epsilon\left[{\left(\frac{r}{\sigma}\right)}^{-12}
- d_{\alpha ,\beta }{\left(\frac{r}{\sigma}\right)}^{-6}\right] \;\; ,
\end{equation}
where $\epsilon$ and $\sigma$ are the units of energy
and distance respectively (for krypton $\sigma$=0.357 nm and 
$\epsilon/k_B$=201.9 K). The parameter $d_{\alpha,\beta}$ is
1 for $\alpha = \beta$=A and 0.5 for $\alpha = \beta $=B.
This choice makes B more gas like.
We adopt the cutoff of $2.5 \sigma$ in the potential.
Our choice of parameters of the system, such as temperature,
densities, and those related to the geometry and potentials of
interactions is ultimately geared towards generating
a situation, by trial and error, in which the densities of fluids
A and B are roughly the same in both the droplet and bubble cases.
Thus the droplet made of fluid A that moves in fluid B is approximately
a symmetric image -- density wise -- of the bubble of fluid B that
moves in fluid A.

In Section 3, we discuss gravity driven motion down the 
$z$-axis of a spherically shaped atomic
solid. We present results on the density and velocity
fields of the surrounding fluid
and confirm the validity of Stokes's law. 
In Section 4, we discuss static properties of droplets
of liquid A surrounded by fluid B and those of B-fluid bubbles
immersed in liquid A. We consider the immiscible case in which
$d_{A,B}$=0.
In Section 5, we give illustrative examples of diffusional
merger of the droplets in an environment unconstrained
by any pore walls.
In Sections 6 and 7, we return to the pore geometry and consider motion
of a single droplet and a single bubble under the influence of a 
gravitational force, $g$. We consider two cases: either the $g$
is applied only to the molecules of the droplet/bubble (Section 6),
or it is applied to all of the molecules of the system (Section 7).
In a true nano-pore, the latter is more realistic. On the other hand,
as a toy model, the former is easier to understand because this
case mimics the situation in which the object moves in a stationary
medium and in which only the immediate neighborhood of the
object is disturbed. In other words, the case of the force acting
only on the object, corresponds conceptually to a motion in a much
larger sized fluid than available in MD simulations.

\section{THE SINGLE FLUID FILLED NANO-PORE}

The geometry of the pore is shown in Figure \ref{wall}. 
The pore is constructed by first generating an fcc-like lattice
in which subsequent sites in the $x,y$, and $z$-directions
differ by $\sigma$.
The wall sites are obtained by cutting out an annulus of the
inner and outer radii of 12$\sigma$ and 15$\sigma$ respectively.
The pore's axial length is 24$\sigma$ and the periodic boundary
conditions are adopted along the $z$-direction.
The primary purpose
of introducing the walls is to pin the fluid at the walls and allow
for dissipation. The periodic boundary conditions in the
$x$ and $y$ directions would not lead to establishment of the terminal speed.
For simplicity, we keep the pore wall atoms frozen at their
crystalline positions.

\vspace*{0.6cm}
\begin{figure}
\epsfxsize=3.in
\centerline{\epsffile{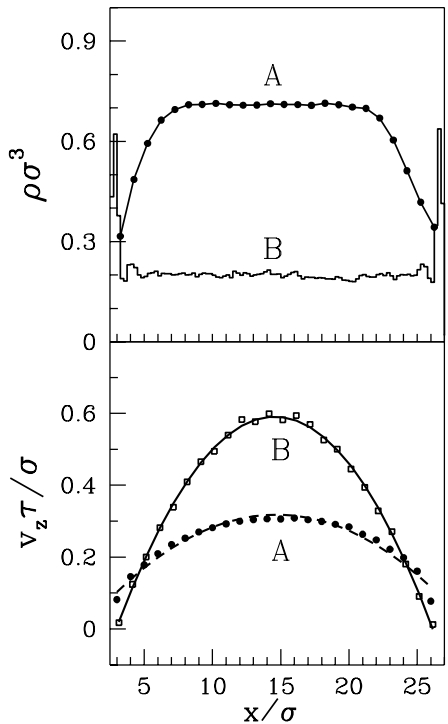}}
\vspace{2.4cm}
\caption{
The top panel shows the
density profiles for fluids A and B, as indicated,
obtained in a vertical cut through the pore. The cut is at $y$=0
and the data are averaged over the axial $z$ direction.
The bottom panel shows velocity profiles for
fluids A and B under gravity driven flow
with $g=0.01\epsilon/\sigma$. The corresponding central densities 
are $\rho=0.2/\sigma^3$ and $\rho=0.75/\sigma^3$.}
\label{dens}
\end{figure}

As is known from the MD studies of channel flows \cite{Banavar,Robbins,Physica}
occuring between two parallel plates, the walls may induce
a layered structure 
in the density profile on the scale of several $\sigma$ in the
neighborhood of the walls. The magnitude of this effect depends on the
strength of the interaction between the fluid and the wall and on the
fluid density.
We adopt the Lennard-Jones interaction for the wall-fluid interaction 
with  $\epsilon$ replaced by $\epsilon _{w,f}=0.7 \epsilon$. Our intention 
here is to provide some pinning
(with a small slip length) and, at the same time, to minimize
existence of the layered structure so that the density profile
is as flat as possible. 
This particular values of $\epsilon _{w,f}$
was obtained by trial and error. Figure \ref{dens}
shows the resulting density profiles for fluids A and B, obtained
with 6554 and 2646 molecules respectively. The densities at the 
central region are correspondingly 0.75 and 0.20 $\sigma ^{-3}$.
For fluid B, there is essentially one layer next to the wall and 
just a remnance of the second atomic layer beyond which
the profile is flat. Larger values of $\epsilon _{w,f}$ would
result in establishment of several layers.
For liquid A, the intermolecular attraction is 
stronger than that due to the wall atoms which results in a
depletion layer near the wall but the flat region in the center
is large enough to accomodate a relatively large sized bubble 
of fluid B, as will be discussed later.
The slip length for fluid B is smaller than for liquid A even though
B is much more rarefied \cite{Physica}. This is due to the different
values of $d_{\alpha , \beta}$ in equation 2 but the same wall-fluid 
interactions for the two fluids.

A characteristic time scale $\tau = \sigma (m/\epsilon )^{1/2}$,
where $m$ is the mass of a fluid molecule (the masses of molecules, 
in fluids A and B are idebtical) corresponds to the
period of oscillations in the Lennard-Jones potential
minimum of fluid A. The integration step was 0.005$\tau$ and the
starting velocity distribution is Maxwellian.
Our thermostatting procedure was based on the
Langevin noise \cite{Grest} which balances the frictional
dissipation. The equation of motion for the $x$-coordinate
of a molecule reads then
\begin{equation}
m\ddot{x}=F_c - \zeta\dot{x}+\Gamma
\end{equation}
where $F_c$ is the force due to other molecules,
$\zeta=0.005m/\tau$ is the coefficient of friction, and 
$\Gamma$ is the Gaussian uncorrelated random force such that
\begin{equation}
\left<\Gamma(t)\Gamma(t')\right>=2\zeta T\delta(t-t') \;\;.
\end{equation}
A similar equation holds for the other coordinates but no Langevin
noise was applied in the $z$-direction in order not to
affect the systematic flows due to gravity, once it is switched on.
The equations of motion are solved by means of the fifth order
predictor-corrector scheme \cite{Tildesley}.
Throughout the paper, we report on the calculations done at
the temperature $T$ of 0.71 $\epsilon/k_B$ at which there is
a sufficient stability of the droplets of the size studied here.
Another reason for this choice of $T$ is that this is one of the several
values chosen in the classic MD studies of droplets by 
Thompson et. al. \cite{Thompson}.

The viscosity of fluids A and B can be obtained from the velocity
profiles, as shown if figure \ref{dens}, by fitting to \cite{batch}
to the parabolic form
\begin{equation}
v=\frac{\rho g}{4\eta}(R_0^2-r^2) \;\;.
\label{parab}
\end{equation}
Here, $R_0$ is the effective radius of the cylinder, 
$\rho$ denotes fluid density, and
$r$ is the polar coordinate.
Our values of $\eta$ for fluids A and B are $2.22 \pm 0.12$
and $0.23\pm0.04 \; \frac{m}{\sigma \tau}$ respectively.

\section{MOTION OF SPHERICAL SOLID GRAINS}

We begin our analysis by considering a gravity driven motion of
a spherically shaped solid object in fluid B. 
Our model of the falling object is an amorphous variant
of a model considered by Vergeles et. al. \cite{Vergeles}.
We first take 454 points
and place them randomly within a spherical volume with the
density of $0.8/\sigma ^3$. The corresponding radius is 5.1$\sigma$.
These points are adopted as tethering centers for molecules
of the same mass as that of the fluid molecules.
The amorphicity is adopted for a more direct comparison
to a liquid droplet which is amorphous but
lacks the stiffness of the solid
The molecules of the solid are attracted to their tethering
centers by a strong harmonic force with the spring constant of 
400 $\epsilon/\sigma ^2$.

The solid atoms in the bulk of the sphere just stay at the tethering
centers but those near the surface, in addition  to the elastic forces,
experience interactions, $u_{s,f}$, with the molecules of the fluid.
We chose $u_{s,f}$ to be
as between the fluid atoms, i.e. of the Lennard-Jones type:
\begin{equation}
u_{sf}(r)=4\epsilon\left[{\left(\frac{r}{\sigma}\right)}^{-12}
- d_{s,f}{\left(\frac{r}{\sigma}\right)}^{-6}\right] \;\; .
\end{equation}
We have considered 3 values of $d_{s,f}$: 0, 0.5, and 1.
The first choice, of a pure repulsion, mimics the interactions
between the A and B fluid molecules  to allow comparisons with 
the droplet and this is the choice we focus on more.
The other choices allow the sphere to drag more fluid
and perturb the environment stronger.

We determine the motion of the solid atoms in two stages.
First, we calculate what are the net force and torque with which
the surrounding fluids acts on the whole solid.
The tethering centers are then translated and rotated, around
the center of mass, according to the values of these quantities.
(We have found no systematic rotation in our studies).
In the second stage, we reevaluate the 
forces on solid's atoms (the elastic contributions come from tethers
whose anchores have been moved) and then evolve the positions
of the atoms according to the standard MD scheme.
Gravitational forces, if any, are applied after an equilibration period,
of order 300$\tau$.

\begin{figure}
\epsfxsize=3.in
\centerline{\epsffile{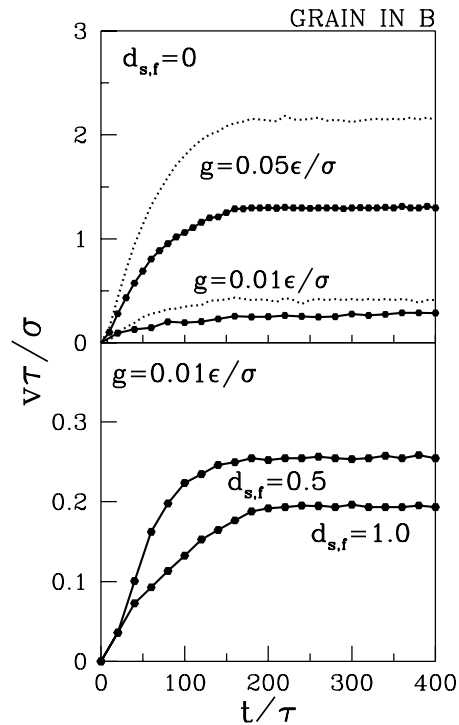}}
\vspace{2.3cm}
\caption{Evolution of the velocity of the center of mass of the spherical 
solid in fluid B for the two different values of $g$ as indicated in 
the figure.
The solid lines correspond to the gravitational force acting
only on the atoms of the solid whereas the dotted lines correspond
to this force being applied to all atoms.
The upper panel is for $d_{s,f}$=0.5 and
the terminal velocities are, top to bottom, 2.148, 1.297, 0.441, 
and 0.266 $\sigma / \tau$.
The lower panel is for $g=0.01 \epsilon /\sigma$ and it compares
two values of $d_{s,f}$: 0.5 and  1. The terminal speeds are
$0.254$ and $0.194 \sigma/\tau$ respectively.
}
\label{rigid}
\end{figure}

Figure \ref{rigid} shows the velocity of the center of mass
as a function of time. The top panel is for $d_{s,f}=0$ and the bottom 
panel for two larger values of $d_{s,f}$.
We observe the phenomenon of saturation --
the center of mass velocity saturates at a terminal velocity $v_t$.
The saturation level is a linear function of $g$. 
and it decreases when $d_{sf}$ is increased. The time
scale to reach the saturation is of order 200$\tau$. From equation
\ref{'stokes'} for the grain we have found value of terminal velocity to
be equal: $v^{'}_{t}=0.203\tau/\sigma$ ($g=0.01\epsilon/ \sigma$) and $v^{'}_{t}=1.019\tau/\sigma$ ($g=0.05\epsilon/ \sigma$).
These values are in good agreement with ones obtained numerically
when gravitational force is acting
only on the atoms of the solid. For other case when his force being applied to all atoms the equation \ref{'stokes'} is not valid.\par
When the gravitational force is applied only to the atoms of
the solid,
the velocity field in the fluid is affected almost exclusively
in the central region of the width that coincides with the
sphere's diameter. 
If the force is applied to all atoms, the terminal speeds are
larger since the fluid itself participates in the motion
-- this becomes a Poiseuille-like flow. \par

\begin{figure}
\epsfxsize=3.in
\centerline{\epsffile{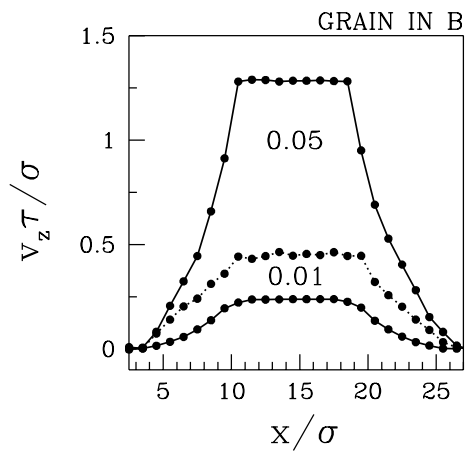}}
\vspace{-1.3cm}
\caption{The velocity profiles for the rigid spherical solid in a reference
frame that moves with the grain and at the center of the grain.
The solid lines correspond to the gravitational force acting
only on the atoms of the solid whereas the dotted lines correspond
to this force being applied to all atoms. 
The values of g, in units of $\epsilon/\sigma$, are indicated.
}
\label{rigid211}
\end{figure}
In Figure \ref{rigid211} profiles of velocity of spherical solid are shown.
These profiles have been obtained in the moving system of coordinates connected
with the center of mass of spherical solid.
The atoms in the central region move rigidly.

\section{STATIC DROPLETS AND BUBBLES}

There have been a number of MD studies of droplets.
Among the more recent entries there are simulations of droplets that form
during a rapid expansion of a liquid \cite{Ashurst} 
and the analysis of the terraced spreading of a 2000-molecule droplet
that lands on an atomic surface \cite{Yang}.
A comprehensive study of liquid droplets in equilibrium with their
own vapor has been done by Thompson et. al. \cite{Thompson}.
They have considered between 41 to 2004 Lennard-Jones molecules of a 
single kind. Still larger droplets have been simulated by 
Powles et. al. \cite{Powles1,Powles2}.
The studies by Thompson et. al. focus  on the quantities related to the
phenomenon of the surface tension. The authors have 
provided a molecular level validation for a) Laplace's equation
for the pressure difference between the inside and outside of a
droplet, b) Kelvin's equation for the radius-dependence of the
vapor pressure, and c) Tolman's equation for the effect of
curvature on the coefficient of the surface tension.
The profiles of the density obtained on radial crossing the gas-liquid 
interface have been found to be well represented by the function:
\begin{equation}
\rho(r)=\frac{1}{2}(\rho_l+\rho_v)-\frac{1}{2}(\rho_l-\rho_v)tanh(2(r-R)/D_s),
\label{th}
\end{equation}
where $D_s$ is a measure of the thickness of the interface,
$R$ is an estimate of the droplet radius,
$\rho_l$ is the density of the liquid in the center of the droplet, 
and $\rho_v$ denotes the density of the vapor far away from the interface.
$D_s$ depends on the gas-liquid interaction and it is
usually of order several $\sigma$.

One theoretical device that was used by Thompson et. al. \cite{Thompson}
was to adopt the geometry of a spherical container that would
scatter the molecules back to the system if they were about to leave it.
When we studied droplets in a 'vacuum' with the periodic
boundary conditions in all directions, we have found that at each 
value of $T$ and of  $\rho _l$ there is a threshold
number of molecules below which the droplet evaporates and occupies
all of the periodic volume.  Above this threshold number the droplet
remains stable and is surrounded by a few vapor molecules.

\begin{figure}
\epsfxsize=3. in
\centerline{\epsffile{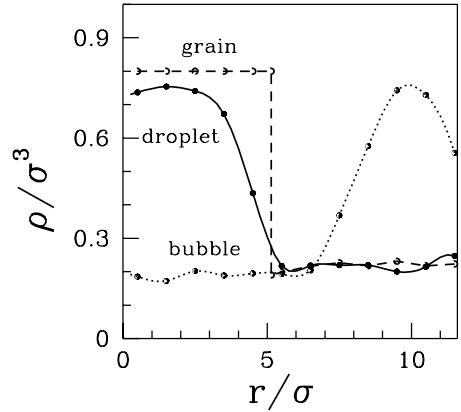}}
\vspace{-0.7cm}
\caption{The profile of the total density corresponding to
a) a droplet of liquid A immersed in fluid B (the solid line),
b) a bubble of fluid B immersed in liquid A (the dotted line),
and c) the spherical solid, of the same number of particles as the droplet,
immersed in fluid B. The molecules occupy the pore space
and the  axial average has been performed -- the profile is
plotted vs. the radius. The droplet data show an enhancement in the
density next to the wall which corresponds to the first monolayer
peak indicated in Figure \ref{dens}
without the axial averaging.
}
\label{prof}
\end{figure}

We have checked that at $T$=0.71$\epsilon /k_B$ a 454-big droplet of
atoms A, with an initial central density of about 0.8$/\sigma ^3$
is stable on its own  -- without any spherical container.
The stability properties are even more assured
when such a droplet is immersed in the immiscible fluid B and this
size of the droplet has been adopted for further studies.
The confining pore walls also add to the stability.
Figure \ref{prof} shows the equilibrated density profile when the droplet is
placed in the pore filled with 2281 molecules of fluid B.
We observe that the droplet is well defined and is not immediately
interacting with the pore walls. Its radius is about the same
as of the spherical solid studied in the previous section --
about 5.1 $\sigma$. The profile at the interface is consistent
with the $tanh$ form given by eq. \ref{th}.\par

In order to generate a stable  bubble, we surround 300 of the
B-type molecules by 6145 A-type molecules. The equilibrated
density profile is shown in Figure \ref{prof}. The corresponding
radius is about 7.4$\sigma$. Thus the bubble is several $\sigma$'s
away from the depletion zone. 
The density (and pressure)  within the bubble is essentially
uniform and it is around 0.2$\sigma ^3$ -- similar to that of the
fluid in which the droplet and the grain move.

\section{COALESCENCE OF DROPLETS}

Before we continue with our analysis of
the drag force phenomena in the pore geometry we pause for some
discussion of coalescence of two initially independent droplets.
Continuum mechanics provides no mechanism for coalescence 
and yet, two droplets placed sufficiently close to each other may
combine simply through atomic diffusion.

\begin{figure}
\epsfxsize=2.7 in
\vspace*{0.5cm}
\hspace*{-1cm}
\centerline{\epsffile{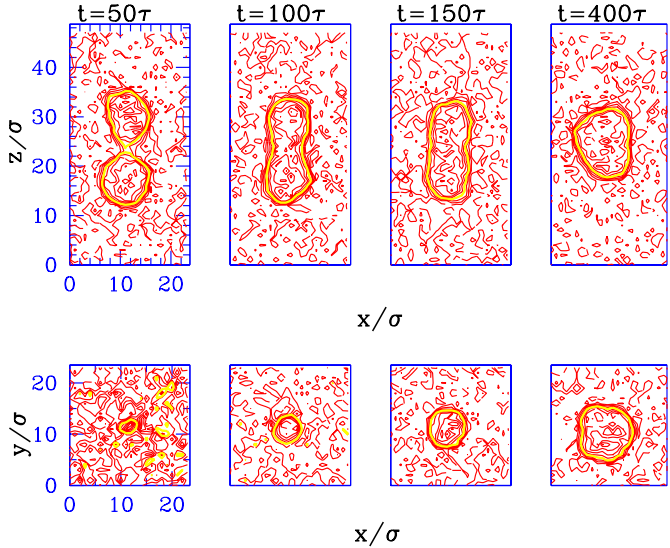}}
\vspace*{0.7cm}
\caption{The density contours of the droplets during coalescence.
The data were time average over periods of 25$\tau$. The top
panels refer to a central cut in the $xz$ plane. The bottom panels
refer to a plane which is half way between the initially separated
droplets and later becomes an equatorial plane in the fused droplet.}
\label{fus}
\end{figure}

In order to show this, we first
consider a volume of size $25\times 25\times 25$ with the periodic
boundary conditions. We form a single A-type droplet of 454 molecules
in the center of the volume and surround it by
3751 molecules of type B (with the density of 0.2 $\sigma ^{-3}$). 
After an equilibration period
of $300 \tau$, we double the volume of the system by placing its
mirror replica in the z-direction. This thus generates two droplets and
the distance between their
the centers of mass is denoted by $R_{12}$.
We have found that when $R_{12}$ is bigger than $14.2 \sigma$,
the two droplets stay as separate and stable entities -- at least during the 
time of 200$\tau$ and and $T=0.71\epsilon /k_B$.
Otherwise they merge diffusively as illustrated in Figure \ref{fus}.
There appears to be no noticeable hydrodynamic velocity field related to the
coalescence.
 
\section{GRAVITY DRIVEN FALL OF DROPLETS AND BUBBLES}

We now return to the problem of the droplet or the bubble motion in 
the pore and consider 
the case in which  $g$ is applied only to the
atoms of the moving object. We ask whether this motion is similar to
that of the rigid amorphous solid.

A hydrodynamic prediction \cite{ryb} for the drag force acting
on a spherical liquid droplet or bubble of radius $R$ is
\begin{equation}
F_d=2\pi R\eta \frac{2\eta+3\eta'}{\eta+\eta'} v\;,
\label{ryb}
\end{equation}
where $\eta$ is the viscosity of the surrounding fluid 
and  $\eta'$ is viscosity of the  fluid of which the moving object is made.
This coincides with the Stokes
law (equation \ref{'stokes'}) in the limit of
$\eta'\rightarrow\infty$. 
In the other limiting 
case of the gas bubble, $\eta'\rightarrow 0$, we obtain:
\begin{equation}
F_d=4\pi R\eta v \;\;.
\label{bub}
\end{equation}
Determination of the terminal speed results from balancing 
the gravitational pull by the drag force combined with the buoyant force
and then \cite{ryb}
\begin{equation}
v_t=\frac{2R^2g(\rho'-\rho)(\eta+\eta')}{3\eta(2\eta+3\eta')}
\end{equation}
where $\rho$ is the fluid density and $\rho'$ is the moving body density.
The sign of $v_t$ agrees with $g$ for droplets but is opposite to $g$
for bubbles. In our MD simulations, the effects of buoyancy are not
incorporated because of the imposition of the periodic boundary conditions
in the axial direction. The corresponding hydrodynamic formula for
$v_t$ corresponds to setting $\rho$ in equation 10 to zero
(and then both a droplet and a bubble move along the $z$ direction).
Due to diffuse density profiles, a more accurate comparison of MD
to the hydrodynamic results is obtained by noting that in the
stationary state $F_d$ should be balanced by $g$ times the number of
molecules in the moving object.  The resulting $v_t$ will be denoted as the
hydrodynamic prediction and the values of viscosity are as
determined by MD in Section II.

\begin{figure}
\epsfxsize=2.8in
\centerline{\epsffile{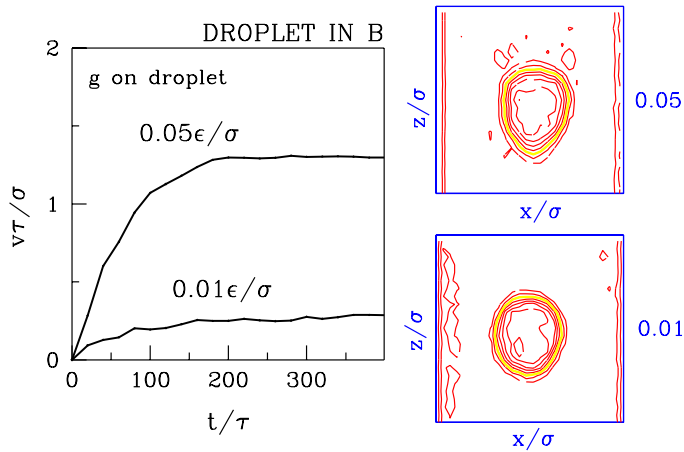}}
\vspace{0.3cm}
\caption{Time evolution of the velocity of mass center of the droplet 
for the two indicated values of $g$. The two panels on the right
show the corresponding density contours  in the stationary state.
The thickest of the solid lines
correponds to a half of the maximum value of the density.}
\label{vd1}
\end{figure}

Results of our MD studies of the droplet motion 
are illustrated in Figure \ref{vd1}.
The left panel shows the time dependence of the velocity of the 
center of mass of the  droplet. The effects of saturation arise
after approximately $100$ 
and  $150\tau$ for $g=0.01$ and $0.05\epsilon/ \sigma$ respectively.
The corresponding $v_{t}$ are 
$0.265\pm0.024$ and $1.293\pm0.021\sigma /\tau$.
The droplet is spherical initially and then it has a radius of $4.4 \sigma$.
The right hand panels
of Figure \ref{vd1} for the density contours in the stationary state
indicate that the droplet
continues to be nearly spherical troughout the motion
if $g$ is sufficiently small.
On the other hand, for $g$=0.05$\epsilon/ \sigma$, there is a noticeable
stretching along the direction of motion (the axial and sideway
radii differ by about 2.7 $\sigma$).

The hydrodynamic prediction for $v_t$ is $0.246 \sigma /\tau$ for
$g$=0.01 if the initial value of $R$. This agrees very well with
the result of MD despite
the perturbing effects of the pore walls. This suggests
that the continuum approach works also for motion of 
microscopic droplets.
Note also that the values of $v_t$ agree very well
with those for the similarly built rigid grain -- the center of mass
moves virtually identically.

The velocity profiles, shown in Figure 8, are quite similar to those
shown in Figure 4 except that there is a parabolic profiling in the
region of the droplet.

\begin{figure}
\epsfxsize=2.8 in
\centerline{\epsffile{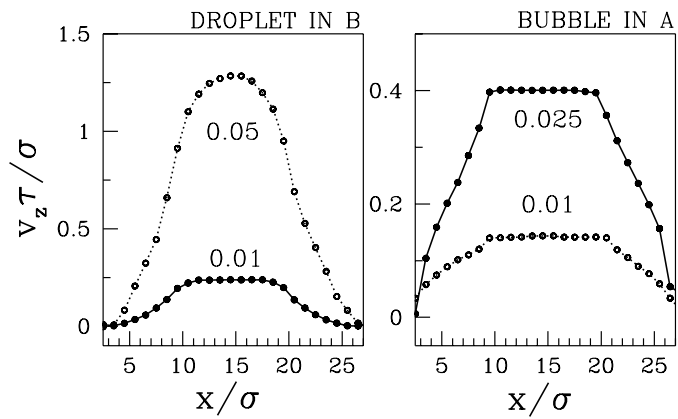}}
\vspace{-0.9cm}
\caption{The left panel shows 
velocity profiles for a droplet  accelerated by force $g$
with the values (in units of $\epsilon / \sigma $) as indicated.
The right hand panel shows velocity profiles for the bubble.
}
\label{pv3}
\end{figure}

Figure 9 shows that the saturation stage of a moving bubble
is reached faster than in the case of the droplet and that
the shape of the bubble gets affected by the motion more
heavily than the droplet. The bubble does not split and moves
as a single entity, at least for $g$ less than $0.025 \epsilon /\sigma$.
The bubble disintegrates for $g$ of order 0.05$\epsilon /\sigma$.
The terminal velocities are $0.154\pm0.07$ and  $0.385\pm0.015\tau/\sigma$
for $g=0.01$ and $0.025\epsilon/ \sigma$ respectively.
The hydrodynamic prediction, however, yields $v_t$ that is
about 10 times smaller. 
This suggests that microscopic bubbles do not move hydrodynamically.
The hydrodynamic behavior might be restored with a much wider pore
but we had no capacity to test this hypothesis.

\begin{figure}
\epsfxsize=2.8 in
\centerline{\epsffile{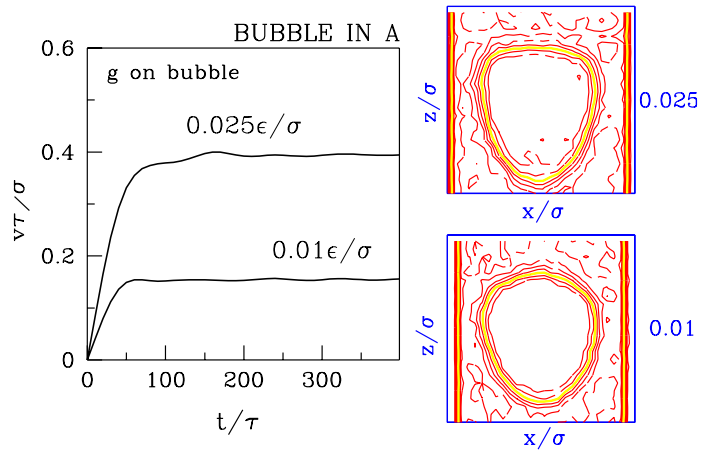}}
\vspace{0.3cm}
\caption{Same as in Figure 7 but for the bubble.}
\end{figure}

Despite the significant shape distortions,
the velocity profiles (Figure 8) for the bubble look very much like
for a falling frozen solid of a comparable density -- there is no
parabolicity in the center.

\section{GRAVITY DRIVEN FLOW WITH DROPLETS OR BUBBLES} 

We now consider the situation in which the gravitational force
is applied to all molecules in the pore space.
This is then two-phase gravity driven flow.

We find that the terminal velocities are now about 2.7 bigger than
the hydrodynamic prediction would yield even though the fluid
right next to the pore wall is almost pinned.
For the droplet, $v_{t}$ is $0.672\pm0.018 $ and $3.272 \pm 0.023\sigma/\tau$ 
for $g$=0.01 and 0.05 $\epsilon/\sigma$ respectively.
For the bubble, $v_t$ is almost the same as when $g$ was applied
only to the molecules of the bubble. For instance, $v_t$ is 
$0.157\pm 0.011 \sigma /\tau$ for $g=0.01 \epsilon /\sigma$.
These results together indicate that when velocity profiling
outside of the moving object become significant the drag force 
deviates from the simple form as given in equation 8.

\begin{figure}
\epsfxsize=2.8 in
\centerline{\epsffile{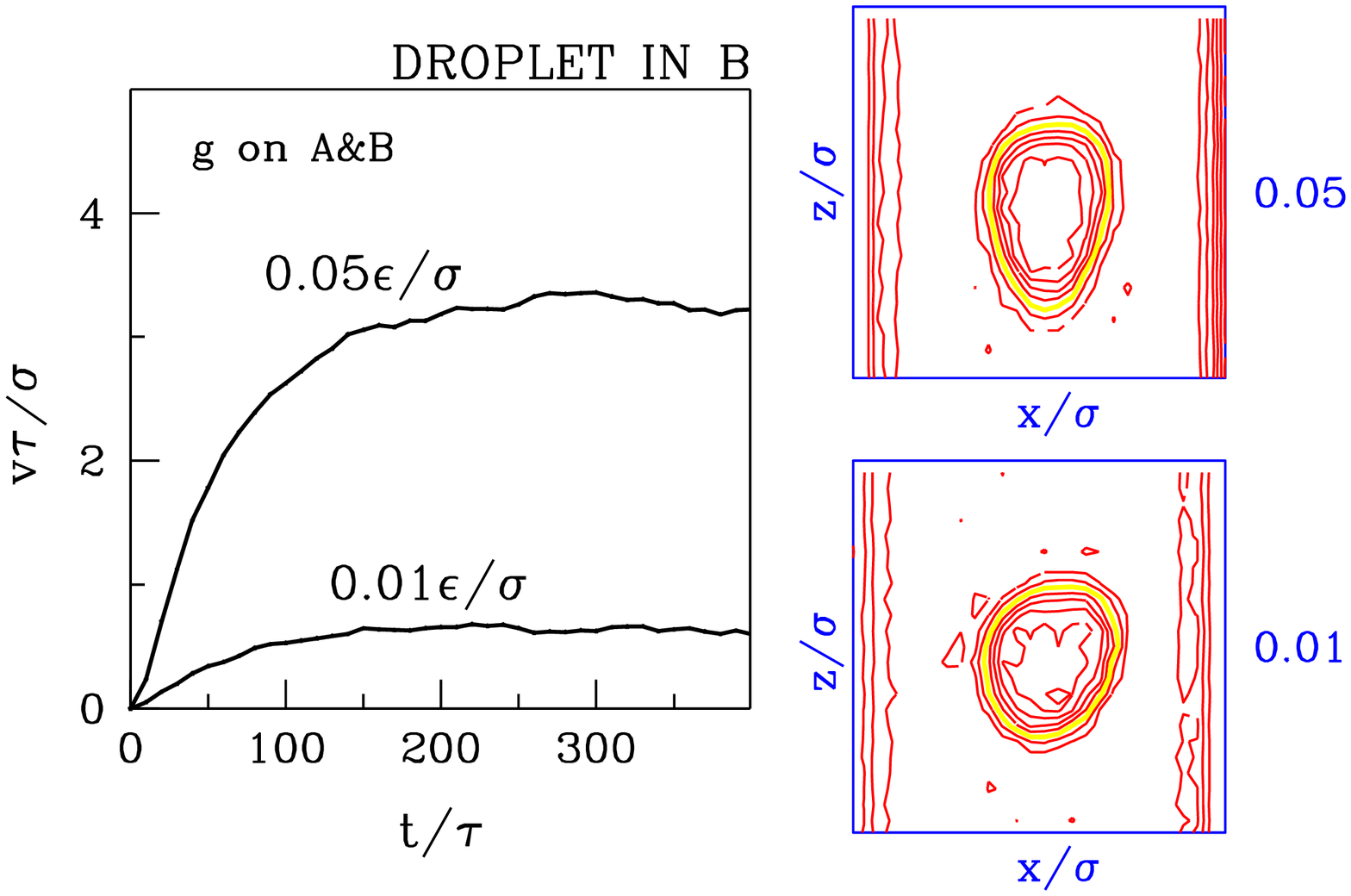}}
\vspace{0.3cm}
\caption{Same as in Figure 7 but for $g$  applied to all atoms within the
poore space.}
\end{figure}

\begin{figure}
\epsfxsize=2.8 in
\centerline{\epsffile{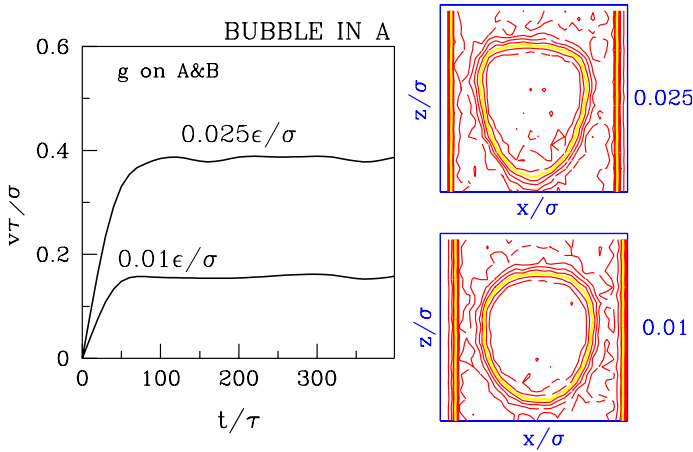}}
\vspace{0.3cm}
\caption{Same as in Figure 9 but for $g$ applied to all atoms within 
the pore space.}
\end{figure}

The density profiles shown if Figures 10 and 11 are close
to those shown in Figures 7 and 9 respectively but the axial
stretching is a bit stronger.

Figure 12 shows the velocity profiles, in the moving frame of reference,
when $g$ is applied to all molecules within the pore and compares it
to the single fluid results as replotted from Figure 2.
Both the droplet and the bubble continue to have velocity
profiles which are flat in the center. When compared to the situation
in which $g$ is applied only to the moving object, as in Figure 8,
the the velocity profile of the droplet is significantly enhanced.
This is consistent with the enhanced values of the terminal velocity.
On the other hand, the velocities around the center of the bubble
are slightly reduced. 
It is interesting to note that, in the central region, the droplet
moves faster than any of the constituting fluids at the same value of
$g$. For the bubble, it is the other way around. This existence of this
effect may depend on the geometry and viscosities involved.
In each case, however, the velocity profile determined in slices
which are vertically away from
the droplet or the bubble agree with the single fluid results of Figure 2.

\begin{figure}
\epsfxsize=2.8 in
\centerline{\epsffile{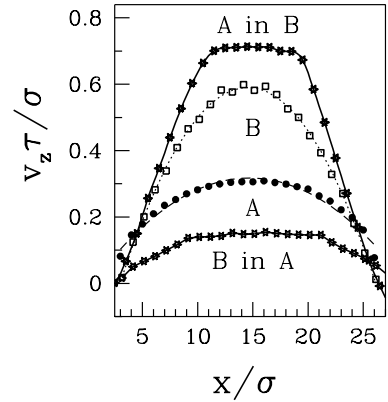}}
\vspace{-0.9cm}
\caption{The velocity profiles when $g$ of 0.01 $\epsilon /\sigma$ is
applied to all atoms withing the pore. The top and bottom curves refer
to the droplet and bubble cases respectively. The data points marked by
A and B are the same as in the lower panel of Figure 2. 
}
\label{vpr1}
\end{figure}

We conclude by noting that the drag force appears to have the simple
hydrodynamic form if the moving object is dense and small compared to the
pore diameter. 
Otherwise, the velocity field is significantly disturbed by the pore and the
drag force becomes more complicated.
It may also be sensitive to the nature of interactions between
the pore wall and fluid atoms.

This research was supported by a grant from the Polish agency KBN (grant
number 2P03B-025-13).



\end{document}